\documentclass[12pt]{article}
\usepackage{latexsym,epsfig}
\usepackage{graphicx}
\usepackage{pifont}
\usepackage{verbatim}
\def\ed{\end{document}}

\usepackage{color}
\def\sk{\smallskip}

\def\bg{\bigskip}
\def\lra{\longrightarrow}
\def\beq{\begin{eqnarray}}
\def\eq{\end{eqnarray}}
\def\beqn{\begin{eqnarray*}}
\def\eqn{\end{eqnarray*}}
\def\nl{\noindent}

\begin{document}
\begin{center}
{\bf \large Vector-  and Scalar-Bilepton 
\sk

Pair Production in Hadron Colliders}
\bg 

{E. Ramirez Barreto}

{Centro de Ci\^encias Naturais e Humanas, UFABC}

{Santo Andr\'e, SP, Brazil}
\vskip .5 cm

{Y. A. Coutinho }

{Instituto de F\'isica, UFRJ, Rio de Janeiro, RJ, Brazil}
\vskip .5 cm

{J. S\'a Borges}

{Instituto de F\'isica, UERJ, Rio de Janeiro, RJ, Brazil}
\end{center}

\begin {abstract}

 We study the double-charged vector-bilepton pair production and double-charged scalar-bilepton pair production {\it via}  
$p + p \longrightarrow Y^{++} + Y^{--} + X$ and $p + p \longrightarrow S_1^{++} + S_1^{--} + X$, where $Y$ and
 $S_1$ are vector and scalar bileptons respectively, in the framework of the minimal version of the 3-3-1 model.
We compute the photon, $Z$, and $Z^\prime$ s-channel contributions for the elementary process of bilepton scalar pair production, and
to keep the correct unitarity behavior for the elementary $q \bar q$ interaction, we include the exotic quark t-channel 
contribution in the vector-bilepton pair production calculation. 
We explore a mass range for $Z^\prime$ and we fix the exotic quark mass within the experimental bounds. In this model, 
the vector-bilepton mass is directly related to $M_{Z^\prime}$ and we consider scalar mass values around the vector-bilepton mass.   

We show that the total cross section for vector-bilepton production is 3 orders of magnitude larger than for scalar pair production 
for  $\sqrt s= 7$ TeV and $14$ TeV and we obtain  the number of events  for the proposed LHC luminosities as a function of the bilepton mass. 
In addition we present some invariant mass and transverse momentum distributions. When comparing these distributions we observe quite
different behavior providing the determination of the bilepton nature. We conclude that one can disentangle
the production rates and that the LHC can be capable of detecting these predicted particles as a signal for new physics.

\end {abstract}

\vfill\eject
\section{Introduction}
The standard model (SM) of the strong and the electroweak interactions  is a very successful theory and from the present year we expect the confirmation of its main predictions. However it was not tested for $\sqrt s > 2$ TeV in hadron colliders, and it is reasonable  to ask if nature will present   new phenomena not described by the  SM. Moreover it is believed that it cannot be the complete theory because it does not explain some theoretical features. This motivates the formulation of many  theoretical 
extensions  or alternatives to the SM. These models for physics beyond the SM  propose, for example, the existence of new particles and their 
predictions  will be tested at the Large Hadron Collider (LHC) and the International Linear Collider (ILC).

In the SM, the vector bosons and the scalar particles do not carry any leptonic or baryonic quantum number, but some extended models predict  
vector bosons and/or scalar particles that couple to two leptons. These particles are known as bileptons and their discovery  would be a signal of
 physics beyond the SM. Another signature of new physics can be the presence of an electrically double-charged particle, which  can decay, for example,  into 
a same sign lepton pair. Many theoretical models  predict particles with these  peculiar properties. For example, 
the doubly charged gauge vector-bilepton  bosons are predicted in the $SU(15)$ grand unified theories \cite{SU15}
and also in models with $SU(3)_c\times SU(3)_L\times U(1)_Y$  gauge symmetry, known as the 3-3-1 model \cite{PIV, FRA,TON}. In the left-right 
symmetric model \cite{LRM} and in the 3-3-1 model the scalarlike bileptons are in multiplets that are introduced for symmetry breaking purposes.
There are composite models, like technicolor \cite{SUS, SUR, SUT}, that introduce nongauge vector bileptons,  and  an extension of the SM has fundamental or composite bileptons \cite{GEO, GEO1}.
\sk

{ In this work, we study the production of bileptons  with  two units of electric charge and from now on we call them bileptons. They  were not observed, but the data extracted from the LEP II experiments can be used to  established limits on the vector-bilepton masses around 100 GeV \cite{GREG}. More exactly, from muonium-antimuonium conversion experiments \cite{MAM, MAM1, MAM2},  the upper bound for the ratio of double-charged vector-bilepton-lepton coupling to its  mass is of order $0.27$ TeV$^{-1}$ with $95\%$ C.L.
\sk

In the case of the double-charged scalar-bilepton, CDF and $D\O$ establish the lower bound for the bilepton mass in the range $110-150$ GeV \cite{MH1, MH2, MH3, MH4} with $95\%$ C.L., from the exclusive bilepton decay into left-handed $e\tau$, $\mu\,\tau$ and $\mu\,\mu$ pairs by considering the left-right model, the SM with a Higgs triplet and the little Higgs model.
\par

The study of vector-bilepton pair production from the hadron collider cannot be  completely
model independent because  the SM Drell-Yan process  violates  unitarity, then requiring  extra $s-$ and $t-$channel contributions from new neutral gauge boson and from exotic fermions 
\cite{DIO}. Having this in mind and in order to focus the LHC physics,  we do elect a model to get correct unitarity behavior for  bilepton pair production in a $p\, p$ collider.
The 3-3-1 model offers the issue for this problem  because in its particle content, together with  simple 
and doubly charged vector gauge bileptons,  it predicts the existence of an extra neutral gauge boson and exotic fermions whose signatures can be manifest themselves at the early stage of LHC operation.

Considering first the gauge sector of the 3-3-1 model, we have explored, in a recent paper, the distributions of dimuons produced at LHC to show a clear signature for the existence of a new  neutral gauge boson, called Z$^\prime$, and compared the invariant mass and angular distributions with those obtained from  other 
models \cite{ELM}. In another application, by considering the $p\, p$ collision, we have calculated the production of a pair of single-charged gauge vector bileptons \cite{EYB}, where the cross section correct unitarity behavior follows from the balance between $Z^\prime$ and exotic quark contributions. There is a similar study for bilepton pair production in $e^+ e^-$ \cite{LON} and a work of the production of just one double-charged vector bilepton \cite{DUT}.

  The production of double-charged Higgs was studied in the left-right symmetric model \cite{RIZ} and in the 3-3-1 model with heavy leptons for  $e^+ e^-$ colliders \cite{TO1}, photon-photon collisions \cite{TO2}, and hadron colliders \cite {TO3, TO4}.

Here we compare the production of a pair of gauge bileptons ($ Y^{\pm \pm}$) with that of a pair of scalar bilpetons  ($S^{\pm \pm}_1$) to evaluate the ratio between the production cross sections. Besides this evaluation we also present some distributions that can reveal the signature associated with the vector- or scalarlike nature of the produced pair and we obtain the number of events for  $\sqrt s = 7$ TeV and  $14$  TeV} at the LHC through  the processes $p + p \lra Y^{++} + Y^{--} + X$ and $p + p \lra {S}^{++}_{1} + {S}^{--}_{1} + X$. 

In Sec. II we review the basic aspects of the minimal version of the 3-3-1 model. In Sec. III we present the tree level calculation of  $q + \bar q 
\lra Y^{++} + Y^{--}$ and $q + \bar q \lra \textit{S}^{++}_{1} + \textit{S}^{--}_{1}$ elementary cross sections as well as the final results, adding some comments.  Finally, in  Sec. IV, we present the conclusions of our work.

\section{Model}

In the 3-3-1 model the electric charge operator is defined  as  
\begin{equation}
Q = T_3 + \beta \ T_8 + X I
\label {beta} 
\end{equation}
\noindent where $T_3$ and $ T_8$ are two of the eight generators satisfying the $SU(3)$ algebra
\begin{equation}
 \left[ T_i\, , T_j\, \right] = i f_{i j k} T_k \quad i,j,k =1, 2,...,8,
\end{equation}
\noindent  $I$ is the unit matrix, and $X$ denotes the $U(1)$ charge.

The electric charge operator determines how the fields are arranged in each representation and depends on the $\beta$ parameter.  
Among the possible choices, $\beta = - \sqrt 3$  \cite{PIV, FRA} corresponds to the minimal version of the model which is used in the present application. 

The lepton content of each generation ($a = 1, 2,  3$) is
\begin{eqnarray}
\psi_{a L} = \left( \nu_{a} \ \ell_a \  \ell^{c}_a \right)_{L}^T\ \sim\left({\bf 1}, {\bf 3}, 0 \right), 
\end{eqnarray} 

\noindent where $\ell^c_a$ is the charge conjugate of the $\ell_a$ ($e$, $\mu$, $\tau$) field. Here the values in the parentheses denote quantum numbers relative to $SU(3)_C$, $SU(3)_L$, and $U(1)_X$ transformations.  

The procedure to cancel model anomalies imposes that quark families must be assigned to different $SU(3)$ representation \cite{CAR}. We elect the left component of the first quark family to be accommodated in $SU(3)_L$ triplet and the second and third families ($m =2, 3$) to belong to the  antitriplet representation, as follows:
\begin{eqnarray}
&& Q_{1 L} =  \left( u_1 \ d_1 \  J_1
\right)_{L}^T \ \sim \left({\bf 3}, {\bf 3}, 2/3 \right) \nonumber \\
&& Q_{m L} =  \left( d_m \  u_m \ j_m
\right)_{L}^T \ \sim \left({\bf 3}, {\bf 3^*}, -1/3 \right)
\end{eqnarray}
\noindent where $a = 1, 2, 3$ and $J_1$, $j_2$, and $j_3$ are exotic quarks with, respectively, $5/3$, $-4/3$, and $-4/3$   units of the positron charge ($e$).
\nl  We will comment about the consequences of our choice in the conclusion section. 

\nl The corresponding right-handed component representations are:
\begin{eqnarray}
&& u_{a R}\ \sim \left({\bf 3}, {\bf 1}, 2/3 \right), \  d_{a R} \ \sim \left({\bf 3}, {\bf 1}, -1/3 \right) \nonumber \\
 && J_{ 1 R}\ \sim \left({\bf 3}, {\bf 1}, 5/3 \right), \  j_{m R} \ \sim \left({\bf 3}, {\bf 1}, -4/3 \right)
\end{eqnarray} 
The gauge bosons  come from the combination of nine fields,  $W^{a}_{\mu}$ ($a=1, 2,...,8$) of $SU(3)_{L}$ and the $U(1)_X$  $B_{\mu}$ field.  We identify the SM $W^{\pm}$ and four additional charged gauge bosons from the combinations:
$$
W_{\mu}^{\pm}= \frac{W_{\mu}^{1} \mp i\,W_{\mu}^{2}}{\sqrt{2}},
V_{\mu}^{\pm}= \frac{W_{\mu}^{4} \mp i\,W_{\mu}^{5}}{\sqrt{2}},
Y_{\mu}^{\pm \pm}= \frac{W_{\mu}^{6} \mp i\,W_{\mu}^{7}}{\sqrt{2}},
$$
where the new charged bosons carry two units of lepton number and are called bileptons.  

In the neutral sector, we define the $\gamma$, $Z$ and the new $Z^\prime$ fields:
\beq
A_{\mu}&=& s_{_W}\, W_{\mu}^{3} \, -\, \sqrt{3}\, s_{_W}\, W_{\mu}^{8}+ \sqrt{1-4\,s^2_{_W}}\, B_{\mu}, \nonumber \\
Z_{\mu}&=& c_{_W}\, W_{\mu}^{3} \,  + \sqrt{3}\, s_{_W} \, t_{_W}\, W_{\mu}^{8} \, -\,  t_{_W} \, \sqrt{1-4\,s^2_{_W}}\, B_{\mu}, \nonumber \\
Z^{\prime}_{\mu}&=& \frac{1}{c_W}\,  \sqrt{1-4\,s^2_{_W}}\, W_{\mu}^{8} \, + \sqrt 3\,  t_{_W}\,  B_{\mu},
\eq
\nl where  $c_{_W}= \cos\theta_{_W}$,  $s_{_W}= \sin \theta_{_W}$, $t_{_W} = s_{_W}/c_{_W}$. 

Usually the neutral physical states $Z_1$ and $Z_2$ are mixtures of $Z$ and $Z^{\prime}$ given  by $$ Z_1 = \cos \, \theta_{mix} \, Z \, -\,  \sin \, \theta_{mix}\, Z^\prime, \quad Z_2 = \sin \, \theta_{mix} \, Z \, +\,  \cos \, \theta_{mix}\, Z^\prime
,$$

\nl where, for a small mixing, $\theta_{mix} \ll 1$, $Z_2$ corresponds to $Z^\prime$, and $Z_1$ to $Z$.
\nl The physical  and symmetry state masses are related to the mixing angle  by 
\beq
\tan^2 \theta_{mix} = \frac{M^2_Z - M^2_{Z_1}} {M^2_{Z_2} - M^2_Z},
\eq
\nl so for a small mixing angle, $M_{Z_1}$ is close to the SM neutral gauge boson mass and $M_{Z_2}$ to the extra neutral gauge boson one.

There are model dependent experimental bounds for this mixing angle \cite{PDG}. In the case of the 3-3-1 model, the mixing becomes small for  $v_\chi >> v_\eta, v_\rho, v_{\sigma_2}$ and vanishes if the $\rho$ and the $\eta$ vacuum expectation values satisfy the relation \cite{DON}
$$ v_\eta^2 \, = \, \frac { 1 \, + 2 \,s_{W}^2 }{ 1 \, - 4 \,s_{W}^2}\,  v_\rho^2,$$ 
\nl independent of the $SU(3)_L$ symmetry breaking scale. 

The minimum Higgs structure necessary for symmetry breaking and
that gives to quarks acceptable masses is:
\beq
\eta =\left(\begin{array}{c} \eta^{0} \\ \eta_{1}^{-} \\ \eta_{2}^{+}
  \end{array}\right)\sim  \left(1,3,0\right)
  ,\,
\rho =\left(\begin{array}{c} \rho^{+} \\ \rho^{0} \\ \rho^{++}
  \end{array}\right) \sim  \left(1,3,1\right)
  ,\, 
\chi =\left(\begin{array}{c} \chi^{-} \\ \chi^{--} \\ \chi^{0}
  \end{array}\right) \sim \left(1,3,-1\right).
\eq
To generate the correct lepton  mass spectrum, one needs a scalar sextet \cite{FOO}
\beq
\Sigma =\left(\begin{array}{llll}
 \sigma^{0}_{1} & \textit{h}^{+}_{2} & \textit{h}^{-}_{1} \\ \textit{h}^{+}_{2} & \textit{H}^{++}_{1} & \sigma^{0}_{2}  \\ \textit{h}^{-}_{1} & \sigma^{0}_{2} & \textit{H}^{--}_{2}  \end{array}\right)
\sim  \left(1,6^*,0\right),
\eq
  
\nl where in parentheses are, respectively, the dimensions of the group representation of $ SU (3)_C $,  $  SU(3)_L$ and the $ U(1)_X$ charges.

The neutral field  of each scalar multiplet develops a nonzero vacuum expectation value ($v_\chi$, $v_\rho$, $v_\eta$, and $v_{\sigma_{2}}$) and the breaking of the 3-3-1 group to the SM is produced {\it via} the following hierarchical pattern:
 $${SU_L(3)\otimes U_X(1)}\stackrel{<v_\chi>}{\longrightarrow}{SU_L(2)\otimes
U_Y(1)}\stackrel{<v_\rho,v_\eta, v_{\sigma_{2}}>}{\longrightarrow}{ U_{em}(1).}$$
The consistency of the model with the SM phenomenology is imposed by fixing a large scale for  $v_\chi$, responsible for giving mass to the exotic particles  ($v_\chi \gg v_\rho, v_\eta, v_{\sigma_{2}}$), with $v_\rho^2 + v_\eta^2 + v^2_{\sigma_{2}}= v_W^2= \left( 246 \right)^2$ GeV$^2$.  

In the minimal version, the relation between $Z^\prime$, $V$, and $Y$ masses \cite{DIO, DNG} is:
\begin{equation}
\frac{M_{V}}{M_{{Z^{\prime}}}} \, =\, \frac {M_{Y}}{ M_{{Z^{\prime}}}} \, =\,  \frac{\sqrt{3 - 12 s_{_W}^2}}{ 2 \, c_{_W}}. 
\end{equation}
This  constraint respects the experimental bounds, and is a consequence of the model, but it  is not often used in the literature. We keep this relation through our calculations.
This ratio is $\simeq 0.3 $ for $s_{_W} =0.23$ \cite{PDG}, and so, in this minimal version of the model,  $Z^{\prime}$ can decay into a bilepton pair.

The relation between the  $SU_L(3)$ and $U_X(1)$ couplings for the minimal version of the  model is
\begin{equation}
\frac {g^{\prime\, 2}}{g^2} =\frac{\sin^2\, \theta_{_W}}{1- 4 \, \sin^2\, \theta_{_W}},
\end{equation}
which fixes $\sin^2 \theta_{_W} < 1/4$, which is a peculiar characteristic of the minimal version of the 3-3-1 model.
\bigskip

The coupling between the gauge bosons and the scalar bosons comes from the gauge invariant kinetic-energy term in the Lagrangian:
\begin{eqnarray}
{\cal L}_{\varphi}^{min}&=& Tr\left( \left( D_{\mu}\varphi\right)^{\dagger}\,\left( D_{\mu}\varphi\right)\right) + V(\varphi),
\label{lt}
\end{eqnarray}

where $\varphi= \eta,\,\rho,\,\chi$  and $\Sigma$. The covariant derivative of the triplet $\varphi= \eta,\,\rho,\,\chi$ is

\begin{eqnarray}
{\cal D}_{\mu}\varphi = \partial_{\mu}\,\varphi - i\,\frac{g}{2}\,{\cal M}_{\mu}\,\varphi-i\,g_{_{X}}\, X_{\varphi}\,B_{\mu} \,\varphi.
\label{lt}
\end{eqnarray}
The covariant derivative of the sextet is
\begin{eqnarray}
{\cal D}_{\mu}\,\Sigma = \partial_{\mu}\,\Sigma - i\,\frac{g}{2}\,\left[{\cal M}_{\mu}\,\Sigma + \Sigma^{T}\,{\cal M_\mu}^{T}\right]\,  -i\,g_{_{X}}\, X_{\Sigma}\,B_{\mu} \,\Sigma
,
\label{lt}
\end{eqnarray}
\noindent with ${\cal M}_{\mu}$ defined as
\beq
{\cal M}_{\mu} =\left(\begin{array}{llll}
 W^{3}_{\mu}+\frac{1}{\sqrt{3}}\,W^{8}_{\mu} & \sqrt{2}\,W^{+}_{\mu} & \sqrt{2}\,V^{-}_{\mu} \\ \sqrt{2}\,W^{-}_{\mu}  & W^{3}_{\mu}-\frac{1}{\sqrt{3}}\,W^{8}_{\mu} & \sqrt{2}\,Y^{--}_{\mu} \\ \sqrt{2}\,V^{+}_{\mu}  & \sqrt{2}\,Y^{++}_{\mu} & -\frac{2}{\sqrt{3}}\,W^{8}_{\mu}  \end{array}\right).
\eq
\nl and   $X_{_\varphi}$ and $X_{\Sigma}$ are the triplet and sextet $U(1)_X$ charges.

The symmetry breaking leads to a shift on the neutral scalar fields, and as a consequence, the  physical states are related to the symmetry states. 
From the more general potential proposed in \cite{LSC}, which involves all triplets and the sextet, we select the terms where the double-charged scalars ($\textit{H}^{++}_{1} $,  $\textit{H}^{++}_{2} $, $\rho^{++}$, and  $\chi^{++}$) are mixed together. From a convenient approximation for the parameters of the double-charged scalar mass matrix we can identify $\chi^{++}$ as a Goldstone state  (to be eaten by  $Y^{++}$). The $\chi^{++}$ decoupling leads to three massive states, the physical state $\textit{S}^{++}_{1}\sim H_1^{++}$ and two physical states from the mixing between  $\rho^{++}$ and $H_2^{++}$. 

 The Lagrangian for the coupling of the sextet to each lepton family is given by

\begin{eqnarray}
{\cal L}_{\Sigma}&=& -\frac{1}{2}\sum_{i, j=1}^3\,  G_{i\,j}\bar \ell^{i}_{L}\,(\ell^{j}_{L})^{c}\,\Sigma^{i\,j} + H.\,c.;
\label{lt}
\end{eqnarray}
\nl
in particular the  interactions involving the double-charged scalar $H_1^{++} \sim \textit{S}^{++}_{1} $ are obtained from:
\begin{eqnarray}
{\cal L}_{\textit{S}_{1}}&=& -\frac{1}{2}  G_{\ell\,\ell}\ \bar \ell^{c}_{\,R}\,\ell_{\,L}\,\textit{S}^{++}_{1} + H.\,c., \qquad \ell = e, \mu , \tau, 
\label{lt}
\end{eqnarray}
\nl where $G_{\ell \ell}\, =\, m_\ell/v_{\sigma_2}$. 

We decide to study this scalar particle production because its main decay mode is into two $\tau$ giving a clear signature for its existence.  The total $S_1^{\pm \pm}$ width is $2\times 10^{-2}$, $2.7\times 10^{-2}$, $4\times 10^{-2}$, and $11\times 10^{-2}$ GeV for $M_{S_1} = 150$, $200$, $300$, and $400$ GeV, by fixing $v_{\sigma_2} = 123$ GeV \cite{LSC}.  

The trilinear gauge couplings (TGC) in this model are obtained from the part of the Lagrangian
describing the self-interactions of gauge fields:
\begin{equation}
{\cal L}_{TGC} = - g\ f_{abc}\ \partial_\mu W_\nu^a \
W^{b \mu}\ W^{c \nu}, \ a, b, c = 1, 2, ..., 8,
\end{equation}
\nl where $f_{abc}$ is the $SU(3)$ antisymmetric structure constant.

Expressing $W^a \ (a = 1, 2, ..., 8)$ in terms of the neutral and double-charged physical
fields, a  straightforward  calculation leads to
\begin{eqnarray}
{\cal L}_{TGC}^{min}&=& - 2 \,i\,g\,s_{_W}\left[ A^\nu (
Y^{--}_{\mu \nu} Y^{++\mu} - Y^{++}_{\mu \nu} Y^{--\mu} ) +
A_{\mu \nu} Y^{--\mu} Y^{++\nu}\right] \nonumber \\
 & & - i\,g\,\frac{1 - 4 \, s_{_W}^2}{2\,c_{_W}} \left[ Z^\nu (
Y^{--}_{\mu \nu} Y^{++\mu} - Y^{++}_{\mu \nu} Y^{--\mu} ) +
Z_{\mu \nu} Y^{--\mu} Y^{++\nu}\right] \nonumber \\
& &  +i\,g\, \frac{\sqrt{3 - 12 \, s_{_W}^2}  }{2\,c_{_W}} \left[ Z'^\nu (
Y^{--}_{\mu \nu} Y^{++\mu} - Y^{++}_{\mu \nu} Y^{--\mu} ) +
Z'_{\mu \nu} Y^{--\mu} Y^{++\nu}\right] \nonumber \\
\label{lt}
\end{eqnarray}
where $Y_{\mu  \nu} \equiv  \partial_\mu Y_\nu -
\partial_\nu  Y_\mu$.
The trilinear couplings used in this paper are shown  in  Table I.
We obtain, from  the expressions (13) and (14), the couplings between the gauge boson and the double-charged scalar $\textit{S}^{++}_{1} $  shown in Table II.
The  charged current interaction of leptons with the vector bilepton is given by
\beq
&&{\cal L}^{CC}=-\frac{g}{\sqrt2}\left(\ell\, ^T\ C \ \gamma^\mu\gamma^5\, \ell
\ {Y^{++}_\mu}\right) + H.\,c.,
\eq
\nl where $C$ is the charge conjugation matrix. 

This double-charged bilepton decays into a pair of equal charged leptons with the same width. They are $1.8,\,  2.3$, and  $2.8$ GeV for $M_Y = 214,\,  271$,  and $325$ GeV, respectively. 
\sk

In the neutral gauge  sector, the  interactions of fermions $\Psi_f$ and bosons are described by the Lagrangian:
\begin{eqnarray}
&&{{\cal L}_{NC}}= \sum_{f} e q_f  \bar
\Psi_f\, \gamma^\mu  \Psi_f A_\mu - \frac{g}{2\, c_{_W}} \bigl\lbrace \bar
\Psi_f\, \gamma^\mu\ (g_{V{_f}} - g_{A_f}\gamma^5)\ \Psi_f\, Z_\mu 
\nonumber \\ &&
+ \bar\Psi_f\, \gamma^\mu\ (g^\prime_{V_f} - g^\prime_{A_i}\gamma^5)\ \Psi_f\, { Z_\mu^\prime}
\bigr\rbrace, 
\end{eqnarray}
\noindent where  $e\,  q_f$ is the fermion electric charge and 
$ g_{V_f}$, $g_{A_f}$, $g^\prime_{V_f}$, and $ g^\prime_{A_f}$ are the fermion vector and axial-vector couplings with $Z$ and $Z^{\prime}$, respectively.

As referred to before, in the 3-3-1 model one quark family must transform  with respect to $SU(3)$  rotations  differently from the other two.  This requirement  manifests itself when we collect the quark currents  in a part with  universal coupling to  $Z^\prime$ similar to the SM and another part corresponding to the  nondiagonal  $Z^\prime$ coupling.
The transformation of these nondiagonal terms, in the mass eigenstates basis,  leads to  the flavor changing neutral Lagrangian
\begin{eqnarray}
{\cal L}_{FCNC}= \frac{g\, c_{_W} }{\sqrt{3 -12 s_{_W}^2}}\ \left(\bar U_L\,\gamma^\mu \, {\cal U}^{\dagger}_L \, B \, {\cal U}_L \, U_L +  \bar D_L \,
 \gamma^\mu \, {\cal V}^{\dagger}_L \, B  {\cal V}_L \, D_L \right) Z^{\prime}_\mu.
\end{eqnarray}
\noindent  where
$$ U_L =  \left( u \,\,\,\, c \,\,\,\,  t
\right)^T_L, \quad  D_L =  \left( d \,\,\,\,  s \,\,\,\, b \right)^T_L \quad {\hbox{and}} \quad  B = {\hbox{diag}}\ \left( 1 \ 0 \ 0\right).$$
The mixing matrices  ${\cal U}$ (for {\it up}-type  quark)  and  ${\cal V}$ (for {\it down}-type quark)   come from the Yukawa Lagrangian and are constrained by the Cabibbo-Kobayashi-Maskawa matrix, as
\begin{equation}
{\cal U}^{\dagger} {\cal V}\,  = \, V_{CKM}.
\end{equation}

By convention in  the SM  {\it up}-type  quark gauge eigenstates are the same as the mass eigenstates, which corresponds to ${\cal U}=I$. This assumption is not valid in the 3-3-1 model because, in accordance to the renormalization group equations, all  matrix elements evolve with  energy and are unstable against radiative corrections, and then  ${\cal U}$ must be $ \not =  I$.  As the Eq. (23) is independent of representation, one is free to choose which quark family representation  must be different from the other two. We recall that here the first family belongs to the $SU(3)$ triplet  representation.  In the next section we will discuss the consequences of our choice.

All universal neutral couplings  diagonal and nondiagonal are presented  in Tables III and IV, respectively.

\begin{table}[h]\label{chihuahua}
\begin{footnotesize}
\begin{center}
\begin{tabular}{||c|c|c|c||}
     \hline
\hline
    & & &          \\ 
  Vertex & $\gamma Y^{++} Y ^{--} $  & $Z Y^{++} Y ^{--} $ & $Z^{\prime} Y^{++} Y ^{--} $\\
    &  & &    \\ \hline
 &  & &  \\ 
Coupling  & $2 e$ &  
   $\displaystyle{e\ \frac{1- 4 \, s_{_W}^2}{2\, s_{_W}\, c_{_W}}}$ 
&     $\displaystyle{- \frac{e}{2\, s_{_W} \, c_{_W}} \, \sqrt {3 - 12
 \, s_{_W}^2}}$ \\
&     & &     \\ \hline
\hline
\end{tabular}
\end{center}
\end{footnotesize}
\caption{Couplings of neutral gauge bosons with  vector-bilepton $Y^{\pm \pm}$.}
\end{table}

\begin{table}[h]\label{pitbull}
\begin{footnotesize}
\begin{center}
\begin{tabular}{||c|c|c|c||}
     \hline
\hline
    & & &          \\ 
  Vertex & $\gamma S_1^{++} S_1^{--} $ & $Z S_1^{++} S_1^{--} $ &  $Z^{\prime} S_1^{++} S_1 ^{--} $ \\
    &   & &       \\ \hline
    & & &      \\
Coupling &
$2 e$ &    $ \displaystyle{\frac{g\, (1-2 \ s^2_{_W}) }{c_{_W}}}$  &
  $\displaystyle{\frac{g \, (1- 4 \, s_{_W}^2)}{6\, c_{_W}}}$  \\
& & &          \\ \hline
\hline
\end{tabular}
\end{center}
\end{footnotesize}
\caption{Couplings of neutral gauge bosons with scalar-bilepton $S_1^{\pm \pm}$.}
\end{table}

\begin{table}[ht]\label{sagui}
\begin{footnotesize}
\begin{center}
\begin{tabular}{||c|c|c||}
     \hline
\hline
&    &          \\ 
&  Vector couplings & Axial-vector couplings    \\
&    &          \\ \hline
\hline
&    &      \\
$Z \bar u_i u_i$ & $\displaystyle{\frac{1}{2}-\frac{4\, s^2_{_W}}{3}}$ &
$\displaystyle{\frac{1}{2}}$  \\
&      &   \\ \hline
\hline
&       &   \\
$Z \bar d_j d_j$  &   $\displaystyle{-\frac{1}{2}+\frac{2\, s^2_{_W}}{3}}$  &
$\displaystyle{-\frac{1}{2}}$  \\
&     &     \\  \hline
\hline
&     &     \\
$Z^{\prime} \bar u_i u_i$ & $\displaystyle{\frac{1-6\, s^2_{_W}-{\cal U}^*_{ii} {\cal U}_{ii} \, c^2_{_W}}{2\,\sqrt{3 -12\, s_{_W}^2}}}$
& $\displaystyle{\frac{1+2\,s^2_{_W}+{\cal U}^*_{ii} {\cal U}_{ii}\, c^2_{_W}}{2\,\sqrt{3 -12\, s_{_W}^2}}}$  \\
&    &    \\ \hline
\hline
&     &        \\
$Z^{\prime} \bar d_j d_j$ &  
$\displaystyle{\frac{1-{\cal V}^*_{jj} {\cal V}_{jj} \, c^2_{_W}}{2\,\sqrt{3  -12\, s_{_W}^2}}}$  &  $\displaystyle{\frac{ 1 - 4\, s_{_W}^2 \, +\, {\cal V}^*_{jj} {\cal V}_{jj} \,c^2_{_W}}{2\,\sqrt{3 -12\, s_{_W}^2}}}$  \\
&      &      \\ \hline
\hline
\end{tabular}
\end{center}
\end{footnotesize}
\caption{The $Z$ and $Z^{\prime}$ vector and axial-vector couplings to quarks ($u_1= u, u_2= c, u_3= t$, and $d_1= d, d_2= s, d_3= b$) in the minimal model and ${\cal U}_{ii}$ and ${\cal V}_{jj}$ are $\cal U$ and $\cal V$ diagonal mixing  matrix elements.}
\end{table}

\begin{table}[ht]\label{perro}
\begin{footnotesize}
\begin{center}
\begin{tabular}{||c|c|c||}
     \hline
\hline
&    &          \\ 
&  Vector couplings & Axial-vector couplings    \\
&    &          \\ \hline
\hline
&    &      \\
$Z^{\prime} \bar c u$ & $\displaystyle{-\frac{{\cal U}^*_{12}\,{\cal U}_{11}\, \cos^2\theta_{_W}}{\sqrt{3-12\,\sin^2\theta_{_W}}}}$ &
$\displaystyle{\frac{{\cal U}^*_{12}\,{\cal U}_{11}\,cos^2\theta_{_W} }{\sqrt{3-12\,\sin^2\theta_{_W}}}}$  \\
&      &   \\ \hline
\hline
&       &   \\
$Z^{\prime} \bar t u$  &   $\displaystyle{-\frac{{\cal U}^*_{13}\,{\cal U}_{11}\, \cos^2\theta_{_W} }{\sqrt{3-12\,\sin^2\theta_{_W}}}}$  &
$\displaystyle{\frac{{\cal U}^*_{13}\,{\cal U}_{11}\,cos^2\theta_{_W} }{\sqrt{3-12\,\sin^2\theta_{_W}}}}$  \\
&     &     \\  \hline
\hline
&     &     \\
$Z^{\prime} \bar t c$ & $\displaystyle{-\frac{{\cal U}^*_{13}\,{\cal U}_{12}\, \cos^2\theta_{_W} }{\sqrt{3-12\,\sin^2\theta_{_W}}}}$
& $\displaystyle{\frac{{\cal U}^*_{13}\,{\cal U}_{12}\,cos^2\theta_{_W} }{\sqrt{3-12\,\sin^2\theta_{_W}}}}$  \\
&    &    \\ \hline
\hline
&     &        \\
$Z^{\prime} \bar d s$ &  
$\displaystyle{-\frac{{\cal V}^*_{12}\,{\cal V}_{11}\,cos^2\theta_{_W} }{\sqrt{3-12\,\sin^2\theta_{_W}}}}$  &  $\displaystyle{\frac{{\cal V}^*_{12}\,{\cal V}_{11}\,cos^2\theta_{_W}}{\sqrt{3-12\,\sin^2\theta_{_W}}}}$  \\
&      &      \\ \hline
\hline
&       &   \\
$Z^{\prime} \bar b d$  &   $\displaystyle{-\frac{{\cal V}^*_{13}\,{\cal V}_{11}\, \cos^2\theta_{_W} }{\sqrt{3-12\,\sin^2\theta_{_W}}}}$  &
$\displaystyle{\frac{{\cal V}^*_{13}\,{\cal V}_{11}\, \cos^2\theta_{_W} }{\sqrt{3-12\,\sin^2\theta_{_W}}}}$  \\
&     &     \\  \hline
\hline
&       &   \\
$Z^{\prime} \bar b s$  &   $\displaystyle{-\frac{{\cal V}^*_{13}\,{\cal V}_{12}\, \cos^2\theta_{_W} }{\sqrt{3-12\,\sin^2\theta_{_W}}}}$  &
$\displaystyle{\frac{{\cal V}^*_{13}\,{\cal V}_{12}\, \cos^2\theta_{_W} }{\sqrt{3-12\,\sin^2\theta_{_W}}}}$  \\
&     &     \\  \hline
\end{tabular}
\end{center}
\end{footnotesize}
\caption{The flavor changing vector and axial-vector couplings to quarks ($u$- and $d$-type ) induced by $Z^{\prime}$ in the minimal model.}
\end{table}

The couplings of ordinary to exotic quarks are driven by the double-charged bilepton as follows:  
\begin{eqnarray}
{\cal L}_{CC}= - \frac{g}{2 \sqrt 2} \bigl \lbrack \overline u \gamma^\mu(1 - \gamma^5)\left({\cal U}_{21} \,\,J_2 + {\cal U}_{31} \,\,J_3\right) + \overline J_1 \gamma^\mu(1 - \gamma^5)\,\, {\cal V}_{11} \,\,d\bigr \rbrack Y^{++}_\mu.
\end{eqnarray}
\noindent where ${\cal U}_{21}$, ${\cal U}_{31}$, and ${\cal V}_{11}$ are mixing matrices elements [Eq. (22)]. From this expression and considering the conservation of the leptonic number, we conclude that  the exotic quarks carry two units of leptonic quantum number and  so they are a class of leptoquarks.   

\newpage
\section{Results}

In this paper we investigate the $SU(3)$ vector-bilepton ($Y^{\pm\pm}$) pair production and the  scalar-bilepton ($S_1^{\pm \pm}$) pair production 
 in the  $p \, p$ collision at LHC.  { The physical scalar-bilepton states emerge from the symmetry breaking mechanism which evolves triplets as well as 
scalar sextet from the Higgs sector of the minimal version of the 3-3-1 model.
In our calculation the bilepton mass is related to the mass of the extra neutral gauge boson $Z^\prime$, also predicted in the model, by  Eq. (10).
 For the $Z^\prime $ mass we adopt some values in accordance with accepted bounds \cite{PDG}: $M_{Z^\prime}$=$800$, $1000$, and $1200$ GeV resulting 
in $M_{Y^{\pm \pm}}= 214$ GeV, $271$, GeV and $325$ GeV, respectively. The corresponding $Z^\prime$ widths are $146$ GeV, $ 180$ GeV, and $218$ GeV. 

Using  the LEP precision measurements data and considering the contribution of the vector bileptons of the 3-3-1 model in the calculation of the 
corrections to $Z\rightarrow \bar b \, b$ decay, one determines the allowed region for the  bilepton mass as a function of the exotic quark mass, $M_Q$. In our 
calculation  we consider $M_Q\, =\, 600$ GeV, respecting the experimental 
bounds \cite{LEQ1, LEQ2, LEQ3} and leading to $M_Y$ from $180$ to $500$ GeV \cite{GON}. 

We consider the elementary process, $ q_i + \bar q_i \lra Y^{++} + Y^{--}$ ($q_i = u, d$), taking into account all contributions: $\gamma$, $Z$, and 
 ${Z^\prime}$ in the $s$ channel and exotic heavy quark contributions  in the $t$ channel. For the scalar pair production in $ q_i + \bar q_i 
\lra S_1^{++} + S_1^{--}$ ($q_i = u, d$) there are only $\gamma$, $Z$, and  ${Z^\prime}$ $s-$channel contributions.  We do not consider gluon fusion 
contributions which were shown to be negligible \cite{TO4}. 
We perform the amplitude algebraic calculation with FORM \cite{VER}, and the details  are presented in the paper \cite{EYB}. 

{Let us consider  $\sqrt s =7$ TeV.  The results shown in Fig. 1 for vector-bilepton pair production correspond to the quark mixing parameters: 
${\cal U}_{21} = 0.0054$, ${\cal U}_{31} = 0.7662$, and ${\cal V}_{11}=0.8667$.  we can note the good behavior of the elementary cross sections that 
present a peak around the ${Z^\prime}$ mass and become broader and smaller as the ${Z^\prime}$ mass increase.

For the analysis of the elementary cross section for scalar-bilepton pair production we consider  $M_{S_1} = 200$ GeV and $400$ GeV.  In Fig. 2
 ($ M_{Z^{\prime}} > 2 M_{S_1}$)  we observe two peaks due to {\it s-}channel interference, which are more clear for the subprocess $\bar d \,  d$;  
one peak is associated to the pair production threshold and another one is near $M_{Z^\prime}$. 
In Fig. 3 ($ M_{Z^{\prime}} \simeq 2 M_{S_1} $)  the second peak disappears, and as expected, the increase of the bilepton mass reduces the 
elementary cross section value.

Next we show our results for the bilepton pair production cross section in $p\, p$ collisions  obtained by integrating the elementary cross section 
weighted by the distribution function for partons in the proton  \cite{CTE} and applying a cut on the pseudorapidity of the final bileptons 
$\vert \eta\vert \le 3$ as well as a cut on the final transverse momentum $p_t > 20$ GeV.

As shown in Fig. 4, the ${Z^\prime}$ mass dependent cross section for vector-bilepton pair production is about 3 orders of magnitude 
larger than the almost constant scalar-bilepton pair production. In contrast with the vector coupling, the 
 scalar-$Z^{\prime}$ coupling does not depend on the $Z^{\prime}$ mass.

Let us consider the invariant mass and transverse momentum distributions.  First, for $ p + p \lra S_1^{++} + S_1^{--} \, + \, X $, we present 
 the invariant mass and transverse momentum distributions  in  Fig. 5  for three values of the $Z^\prime$ mass and $M_{S_1} = 200$ GeV. In the 
invariant mass distribution we find a broad peak corresponding to the $Z^\prime$ mass  while in the transverse momentum distribution this shape appears 
close to $M_{Z^{\prime}}/2$.

In Fig. 6 we display the comparison between the normalized transverse momentum distributions by fixing  $M_{Y} = 271 $ GeV and $M_{S_1}=$ 200 GeV. 
We note  different shapes: the vector bileptons are mainly produced with high transverse momenta while the scalar pair production occurs at lower
 momenta. This is also a nice signature to distinguish the bilepton nature. On the other hand, the normalized scalar and vector rapidity distributions
 are flat and similar and so they do not allow one to disentangle the produced particles.

To complete our analysis we consider $\sqrt s \, = \, 14$ TeV for $M_{Z^\prime} = 1000$ GeV. In Fig. 7 we display the comparison between the 
$Z^\prime$ dependent total cross sections and we realize that the vector-bilepton pair production is again about 3 orders of magnitude larger than for a pair
 of scalars.
 
Next we compare, in Fig. 8, the normalized invariant mass distribution for $M_{Y}= 271$ GeV and for two values of the $S_1$ mass ($150$ and $300$ GeV).
 We observe that the distributions are quite different: for the vector bilepton just one peak appears at $M_{Z^\prime}$, whereas, for the scalar one, 
there is another  enhancement for the small scalar-bilepton mass pair.

In Fig. 9, the transverse momentum  distributions show that it is possible to disentangle the produced bileptons: exactly as for $\sqrt s \, =\,  7$ TeV, one vector bilepton is produced 
with higher momentum than one of the scalar bilepton.

We show, in  Fig. 10, the number of events as a function of the bilepton mass. The expected number of vector-bilepton pairs produced by year is in the range $10^2$ to $10^4$ for $\sqrt 
s \, = \, 7$  TeV and $10^5$ to $10^6$ for $14$ TeV, 
corresponding to  $1$ and $100$ fb$^{-1}$ integrated luminosities, respectively. The expected number of scalar-bilepton pairs is about 3 orders of magnitude smaller.

Finally we  explore the consequences of  $Z-Z^\prime$ mixing, expressed by Eq. (7).    
 The data extracted from LEP, SLC and Tevatron experiments  indicate a range $-1.3 \times 10^{-3} \le \theta_{mix} \le 0.6 \times 10^{-3}$ for the left-right symmetric model and $-1.6 \times 10^{-3} \le \theta_{mix} \le 0.6 \times 10^{-3}$ for the $E_\chi$ model \cite{PDG}. We present the explicit dependence of the bilepton production  cross section with the mixing angle for the minimal version of the 3-3-1 model, used in this work. In order to do that, we adopt very large values for this parameter and we show  in the Fig. 11 (top and bottom) our findings. First, for 
vector-bilepton production, we do not observe any dependence on $\theta_{mix}$ for  $\sqrt s = 7$ TeV and $\sqrt s = 14$ TeV. Next, for the production of scalars, we note a weak dependence on $\theta_{mix}$ when this mixing is completely out of thw expected experimental window. On the other hand, as the 3-3-1 model reproduces the low energy SM phenomenology, we assume that it cannot modify substantially, for instance, the $\rho$ parameter. As shown in \cite{ERL}, the precision electroweak data strongly constrain the $Z$-$Z^\prime$ mixing angle at low energy. We expect that this conclusion is also valid for the present model, and  so we do not introduce any $Z-Z^\prime$ mixing in our calculation. 

\section{Conclusions}

In this paper we focus on the scalar- and vector-bilepton pair production in the $p p$ collision at LHC. These particles are predicted in some extensions of the SM and, in particular, in the 3-3-1 model used in the present paper. For the elementary Drell-Yan process  we consider the contributions of $\gamma$, $Z$,  and $Z^\prime$ in the s channel for the scalars' production, whereas for the vector-bilepton pair production we add the t-channel exchange of the exotic quarks $J_1$, $j_2$, and $j_3$. We do not consider the very small gluon fusion contribution to these processes.

 We take into account the mixing of t-channel quark mass eigenstates originating  from the Yukawa coupling, 
 and obtain a set of mixing parameters allowing for good behavior of the elementary cross section.  The resulting parameters are related to our particular choice for $SU(3)_L$  family representation. This result does not exclude  any other choice for quark 
representation.

 We show that the cross section for vector pair production is about 3 orders of magnitude larger than that for scalar $S_1$. To reinforce the possibility to determine the nature of the produced bilepton, we obtain normalized invariant mass and transverse momentum distributions shown in  Figs. 9 and 10. 

Let us mention that the production of this doubly charged scalar bilepton was not 
yet studied in the literature; however, there is an application of the 3-3-1 model with heavy leptons which consider the production of a double-charged Higgs belonging to a triplet $SU(3)_L$ representation, with different decay channels. We recall that the scalar considered here decays mainly into a pair of $\tau$ leptons,  leading to a very clear signature when compared 
with that from the  background corresponding to two $Z$, two $W$, or one $Z$. 

We also show the very small dependence of the total cross section on $Z-Z^\prime$ mixing. We are sure that no mixing is to be considered in our calculation motivated by the experimental results from LEP, SLC, and Tevatron and the low energy electroweak precision data that suggest a very small mixing.

Finally, from  the expected number of events we conclude that it is possible to confirm the very existence of a new physics at 
the first stage of the LHC operation. 

\begin{figure} 
\begin{center}
\includegraphics[height=.4\textheight]{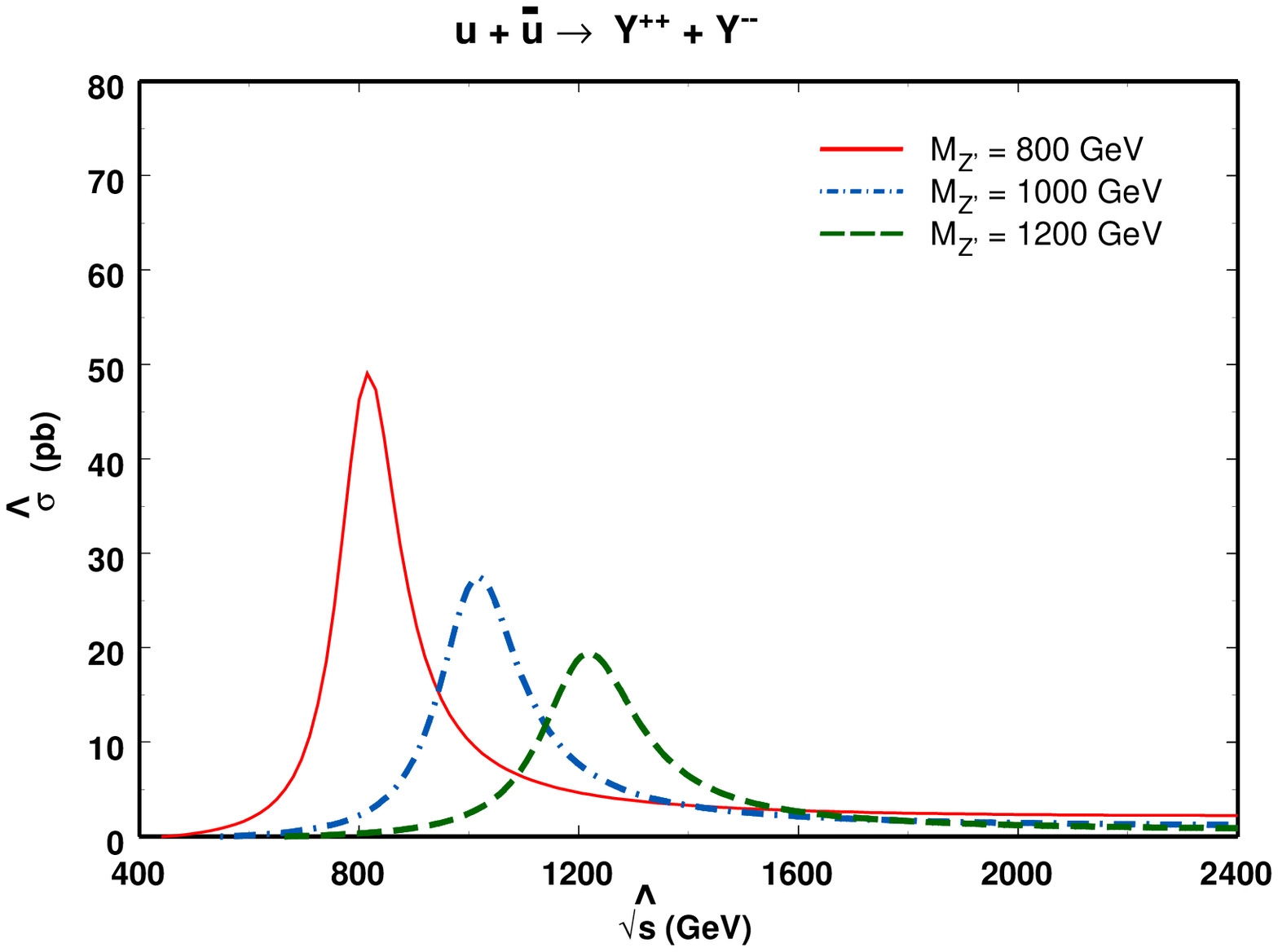}
\hskip 1. cm \includegraphics[height=.4\textheight]{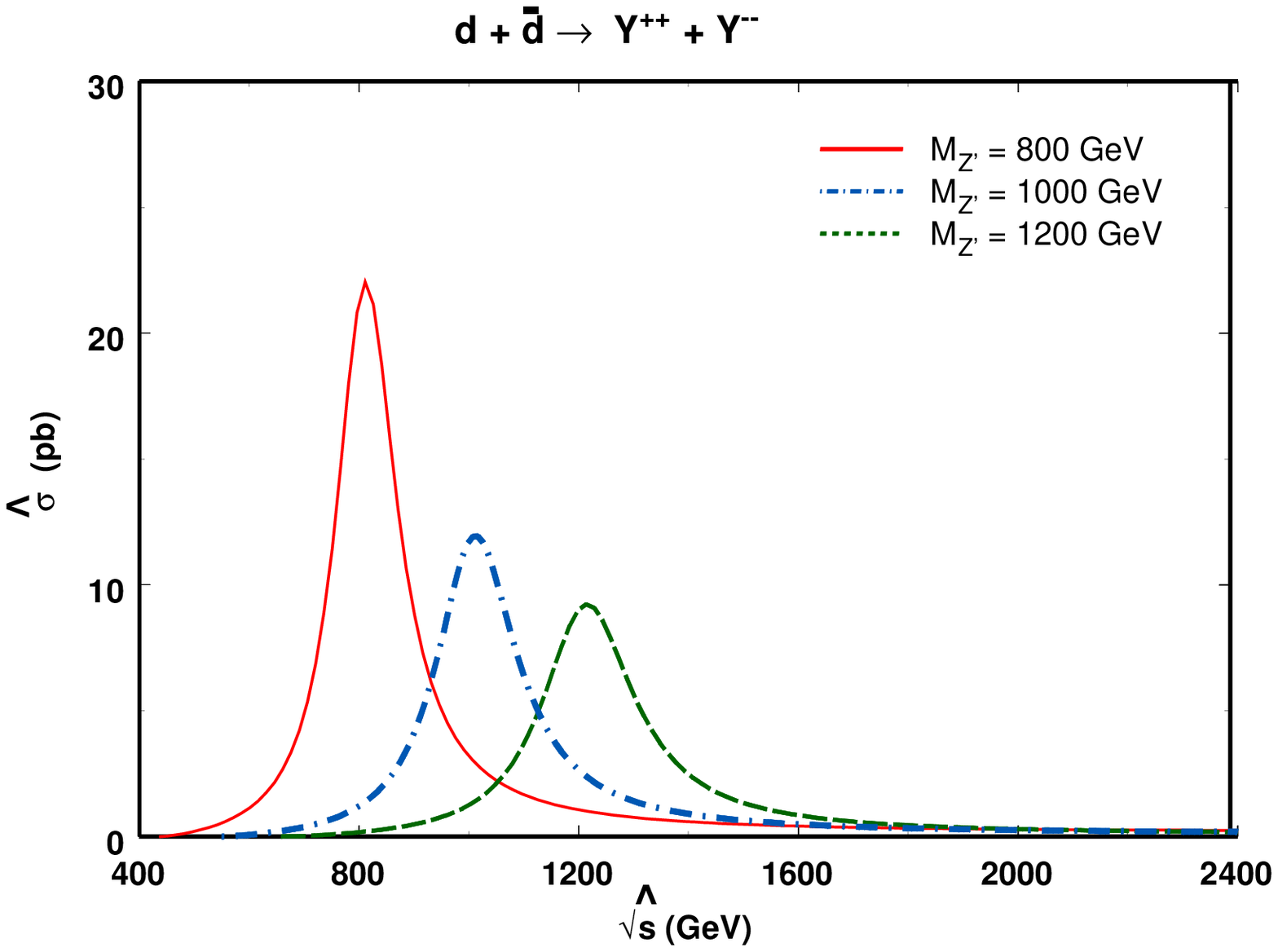}
\caption{ The cross section  for the elementary  process $u + \bar u \lra Y^{++} + Y^{--}$ (top) and 
  $ d + \bar d \lra Y^{++} + Y^{--}$ (bottom) for $M_{Z^\prime}= 800$ GeV, $1000$ GeV, and $1200$ GeV.}
\end{center}
\end{figure}

\begin{figure} 
\begin{center}
\includegraphics[height=.4\textheight]{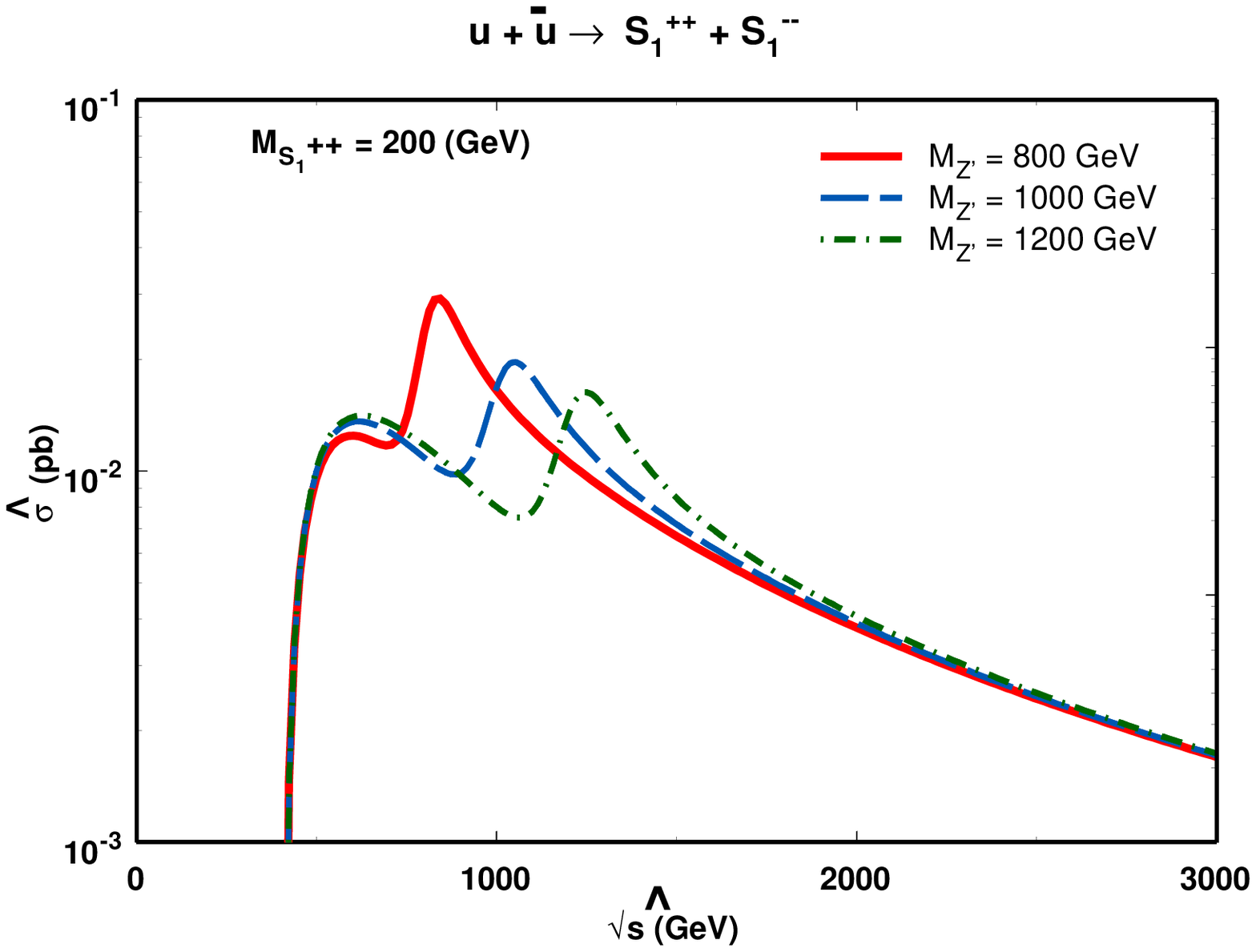}
\hskip 1. cm \includegraphics[height=.4\textheight]{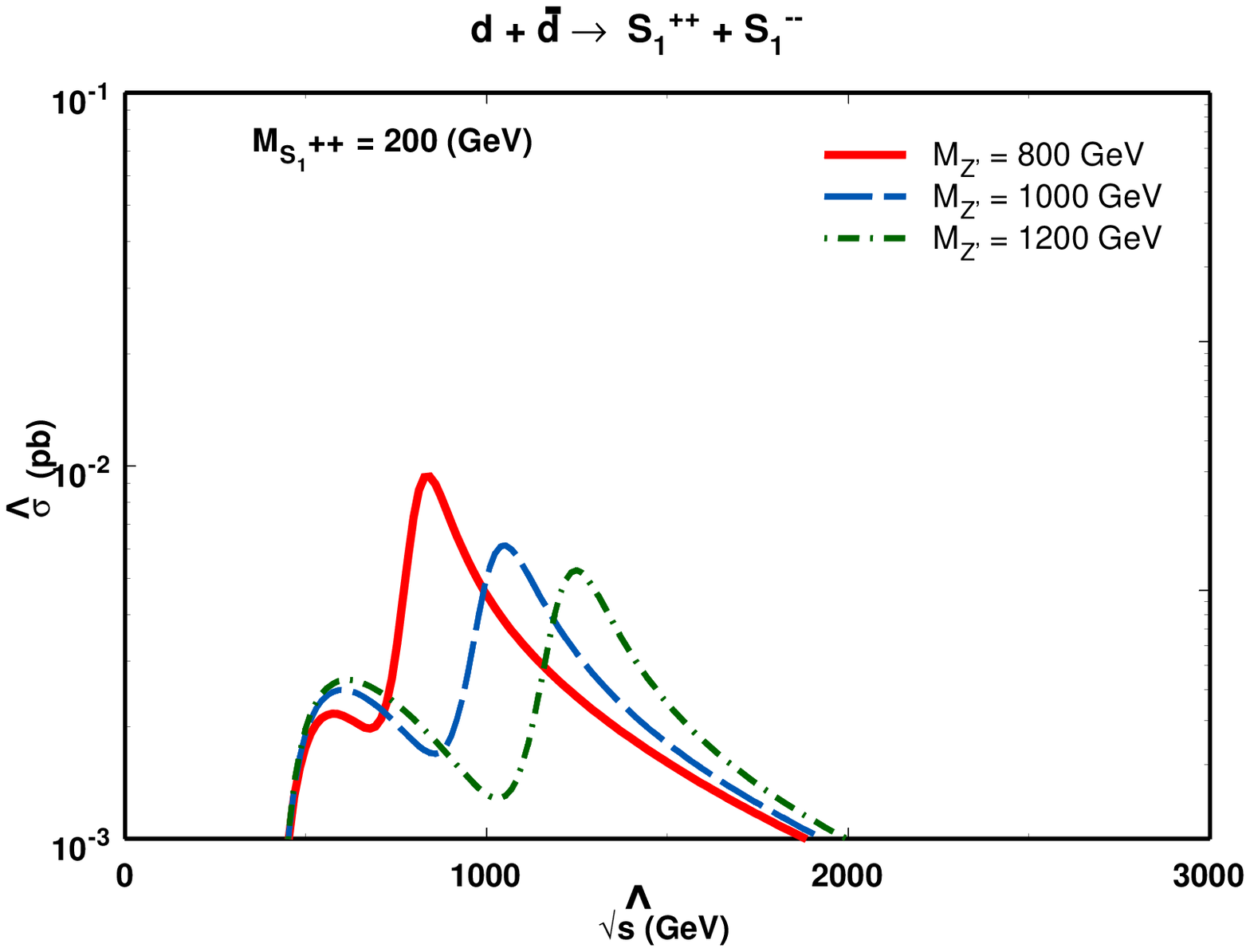}
\caption{The cross section  for the elementary  process $ u + \bar u \lra S_1^{++} + S_1^{--}$ (top) and 
 $ d + \bar d \lra S_1^{++} + S_1^{--}$ (bottom) for $M_{Z^\prime}= 800$ GeV, $1000$ GeV, and $1200$ GeV and $M_{S_1}= 200 $ GeV.}
\end{center}
\end{figure}

\begin{figure} 
\begin{center}
\includegraphics[height=.4\textheight]{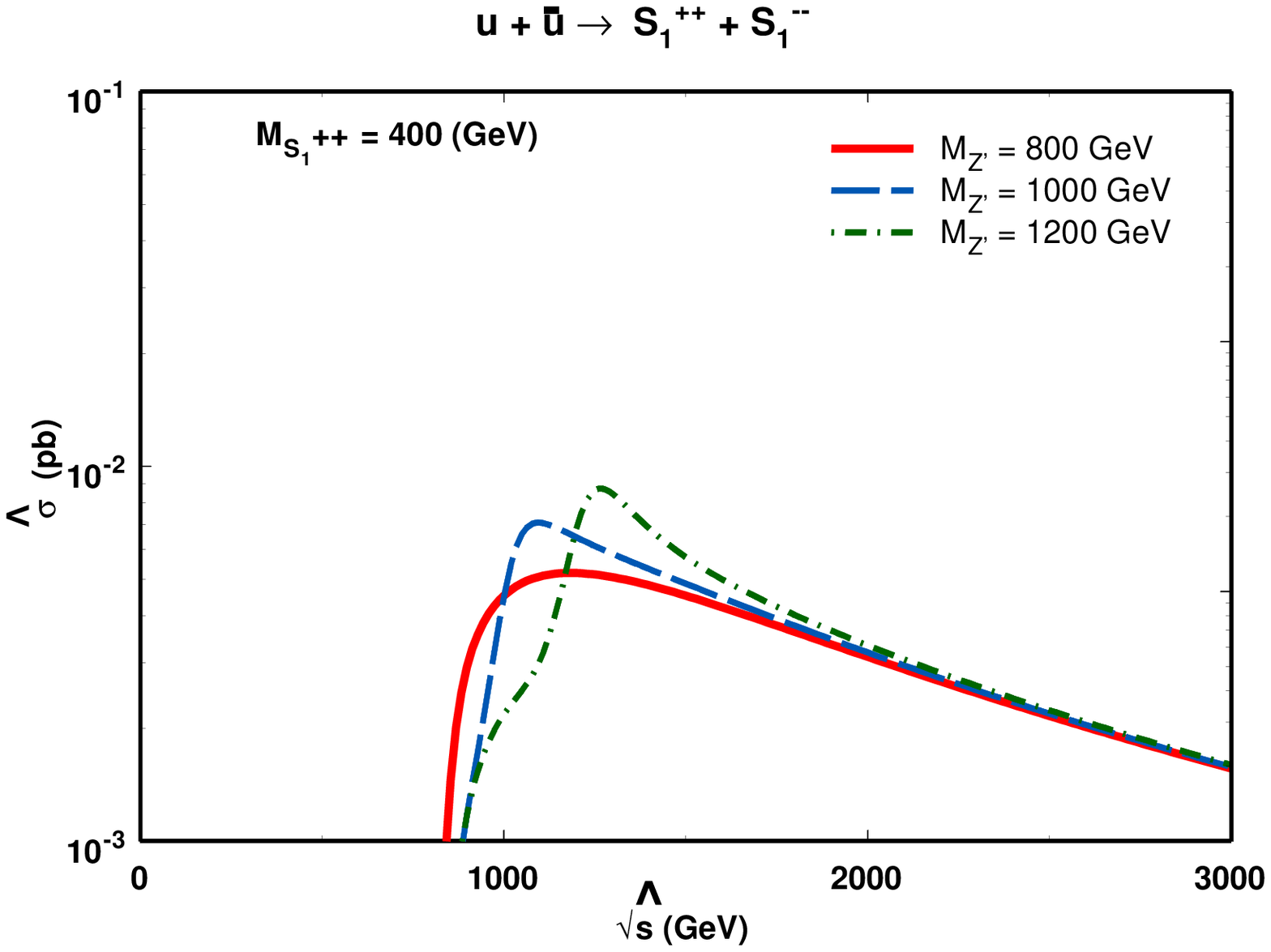}
\hskip 1. cm \includegraphics[height=.4\textheight]{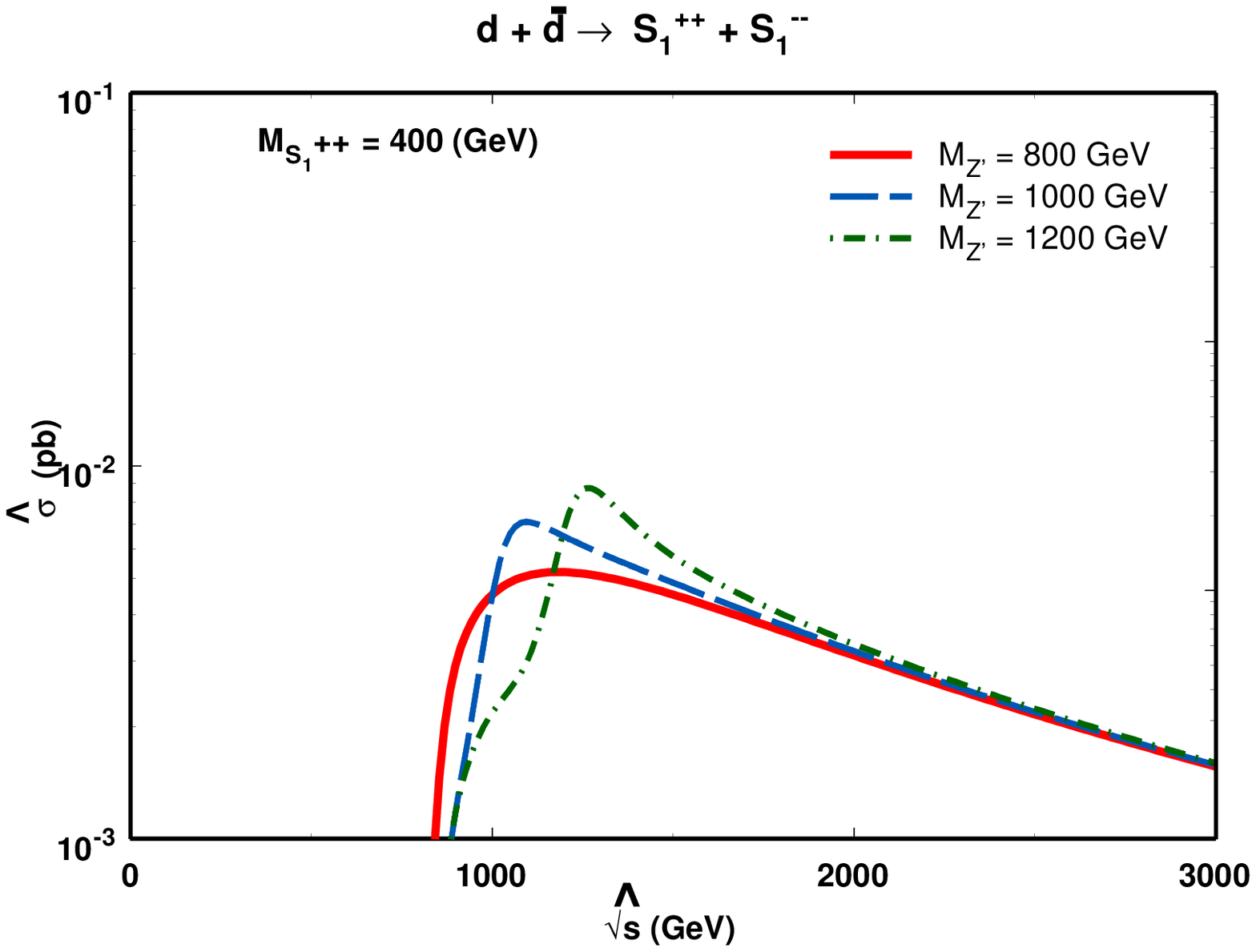} 
\caption{The cross section  for the elementary  process $ u + \bar u \lra S_1^{++} + S_1^{--}$ (top) and  $ d + \bar d \lra S_1^{++} + 
S_1^{--}$ (bottom) for $M_{Z^\prime}= 800$ GeV, $1000$ GeV, and $1200$ GeV and $M_{S_1}= 400 $ GeV.}
\end{center}
\end{figure}

\begin{figure} 
\begin{center}
\includegraphics[height=.4\textheight]{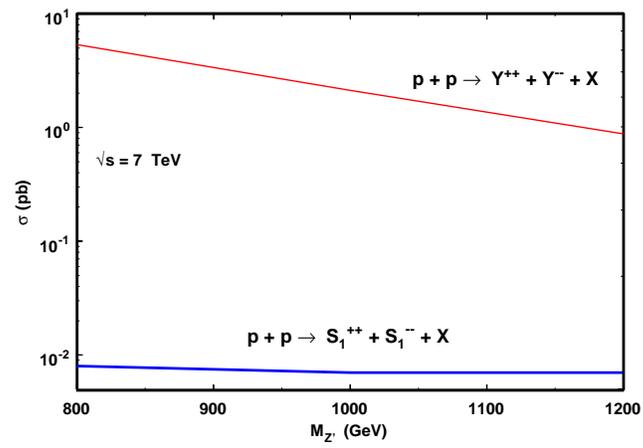}
\caption{The total cross section for the processes $p + p \lra Y^{++} + Y^{--} + X$ (blue solide line) and $p +  p \lra S_1^{++} + S_1^{--} + X $ (red dashed line) for $\sqrt s = 7$ TeV.}
\end{center}
\end{figure}

\begin{figure} 
\begin{center}
\includegraphics[height=.4\textheight]{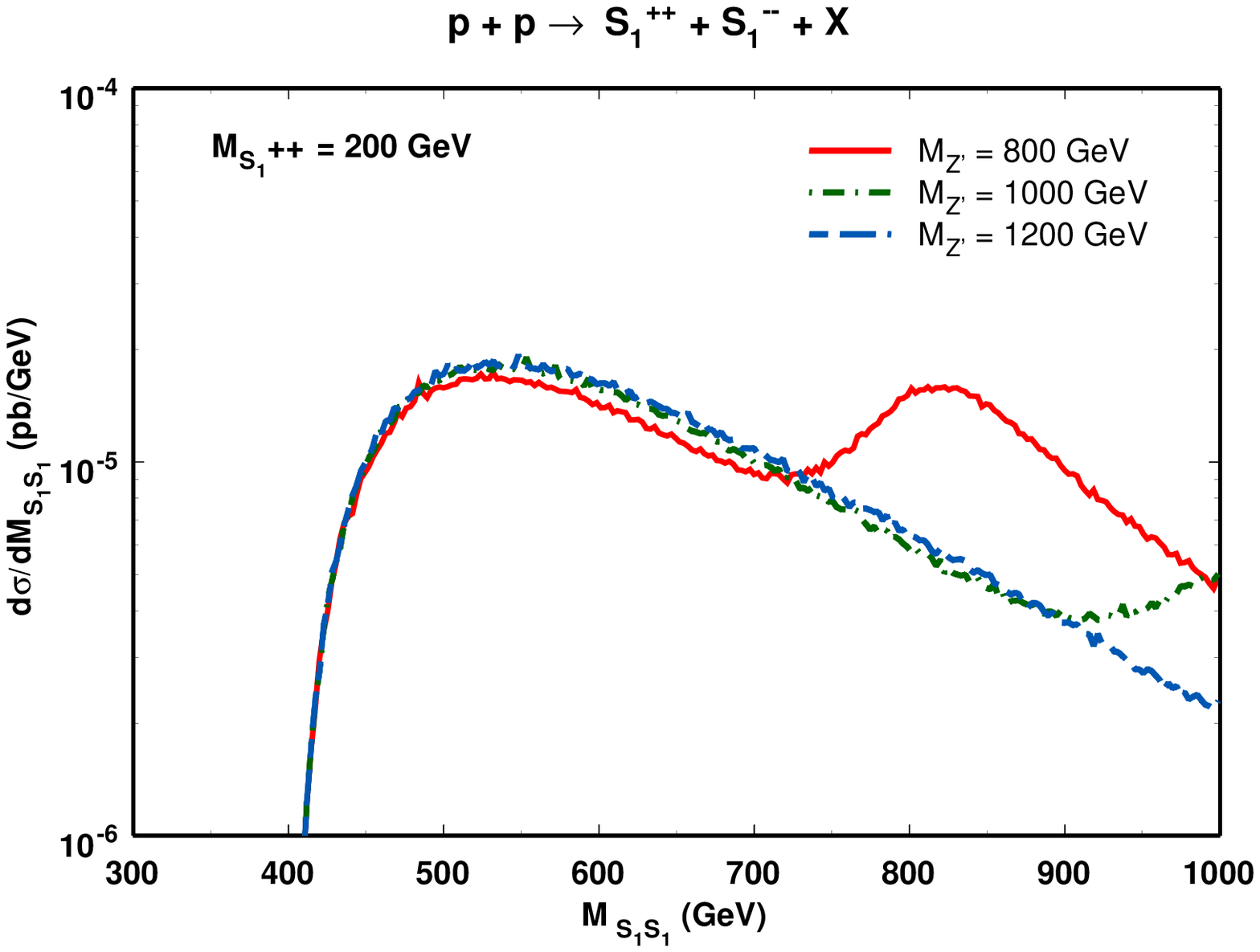}
\hskip 1. cm \includegraphics[height=.4\textheight]{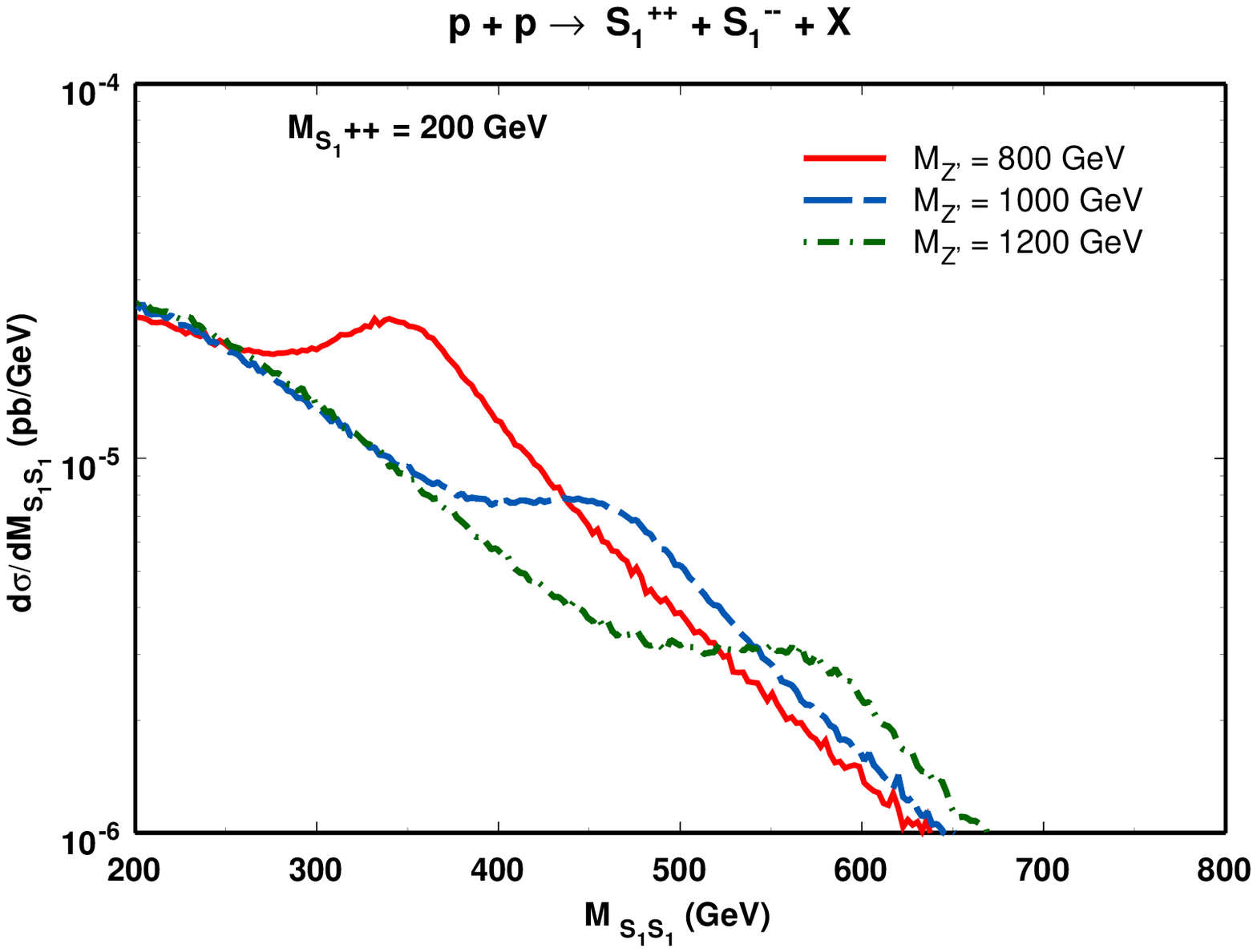}
\caption{The invariant mass distribution for the process $ p +  p \lra S_1^{++} + S_1^{--} + X $ (top) and the momentum transverse 
distribution (bottom) for $M_{Z^\prime}= 800$ GeV, $1000$ GeV, and $1200$ GeV and $M_{S_1}= 200 $ GeV and for $\sqrt s = 7$ TeV.}
\end{center}
\end{figure}

\begin{figure} 
\begin{center}
\includegraphics[height=.4\textheight]{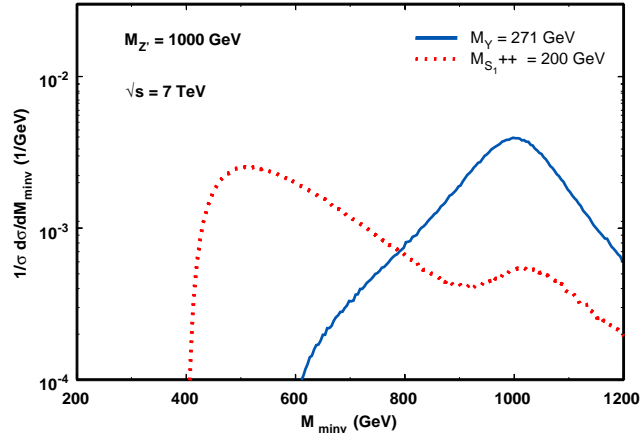}
\hskip 1. cm \includegraphics[height=.4\textheight]{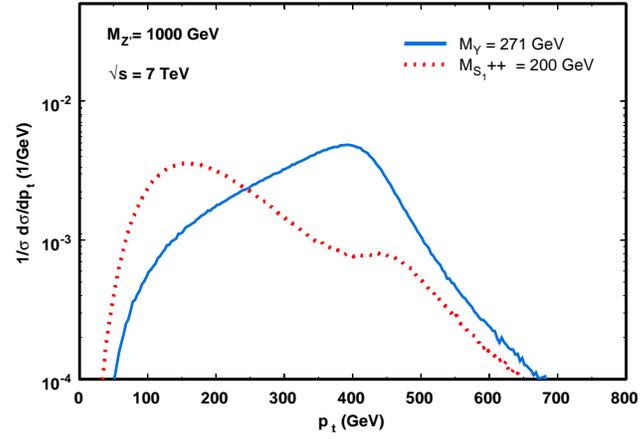}
\caption{The normalized invariant mass distribution (top) and normalized momentum transverse distribution (bottom) for $\sqrt s = 7$ TeV.}
\end{center}
\end{figure}

\begin{figure} 
\begin{center}
\includegraphics[height=.4\textheight]{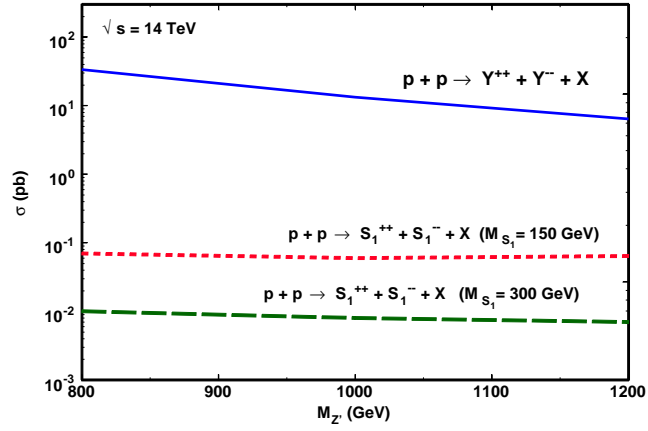}
\caption{The total cross section for $ p + p \lra Y^{++} + Y^{--} + X $ (blue solid line) and $ p +  p \lra S_1^{++} + S_1^{--} + X $ (red short-dashed and green long-dashed line) as 
a function of $M_{Z^\prime}$  ($\sqrt s = 14$ TeV).}
\end{center}
\end{figure}

\begin{figure} 
\begin{center}
\includegraphics[height=.4\textheight]{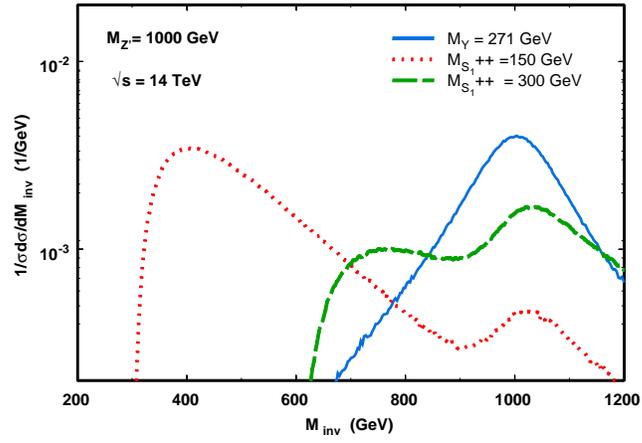}
\caption{The normalized invariant mass distribution for $ p + p \lra Y^{++} + Y^{--} + X $ and $ p +  p \lra S_1^{++} + S_1^{--} + X $ 
 ($\sqrt s = 14$ TeV).}
\end{center}
\end{figure}

\begin{figure} 
\begin{center}
\includegraphics[height=.4\textheight]{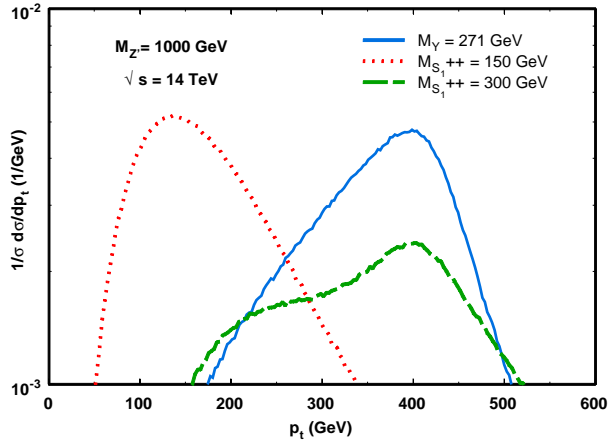}
\caption{The normalized momentum transverse distribution for $ p + p \lra Y^{++} + Y^{--} + X $ and $ p +  p \lra S_1^{++} + S_1^{--} + X $ 
 ($\sqrt s = 14$ TeV).}
\end{center}
\end{figure}

\begin{figure} 
\begin{center}
\includegraphics[height=.4\textheight]{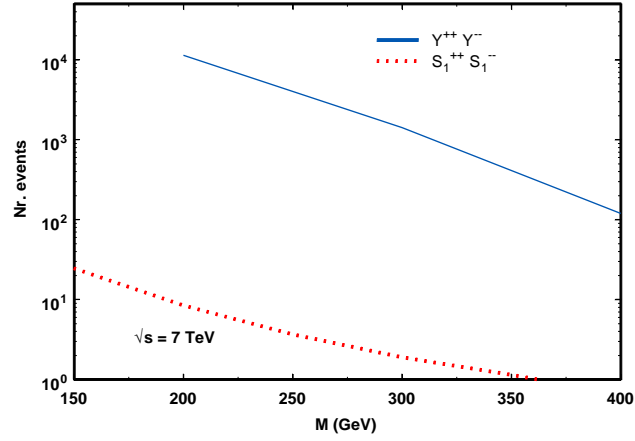}
\hskip 1. cm \includegraphics[height=.4\textheight]{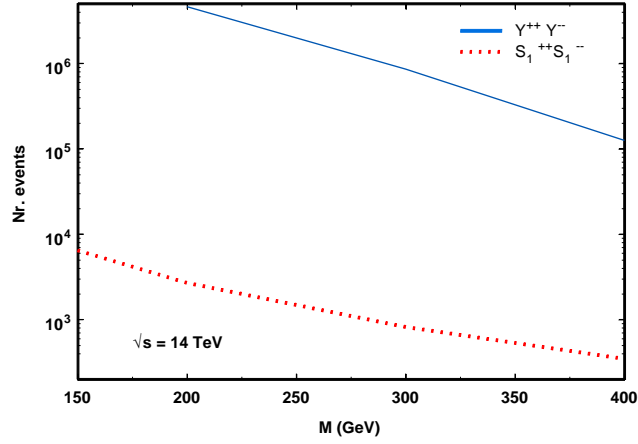}
\caption{Number of events for $ p + p \lra Y^{++} + Y^{--} + X $ (blue solid line) and $ p +  p \lra S_1^{++} + S_1^{--} + X $ (red dotted line) as 
a function of the bilepton mass. For  $\sqrt s = 7$ TeV (${\cal L} = 1$ fb$^{-1}$) (top) and $\sqrt s = 14$ TeV (${\cal L} = 100$ fb$^{-1}$) (bottom).}
\end{center}
\end{figure}

\begin{figure} 
\begin{center}
\includegraphics[height=.4\textheight]{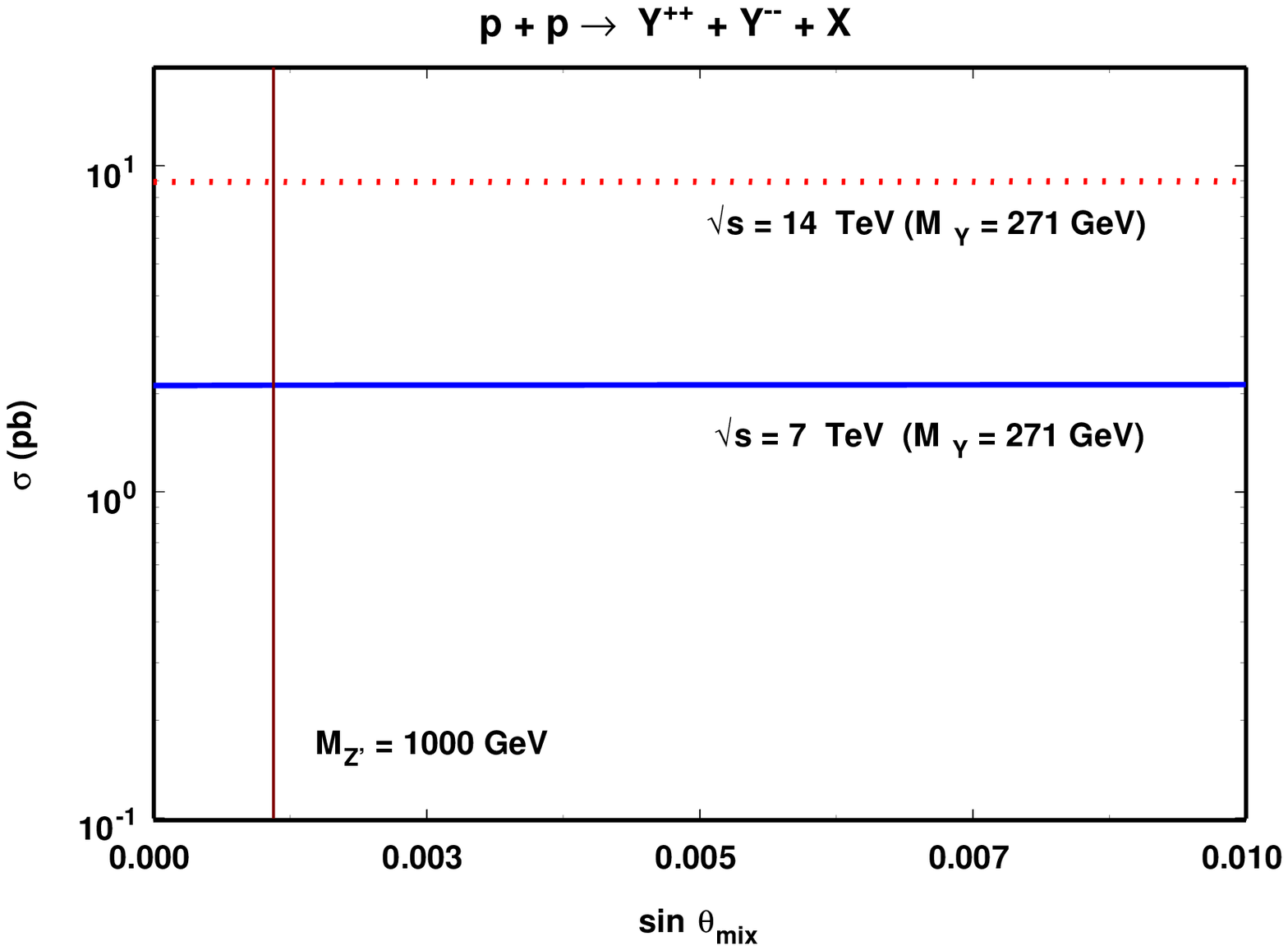}
\hskip 1. cm \includegraphics[height=.4\textheight]{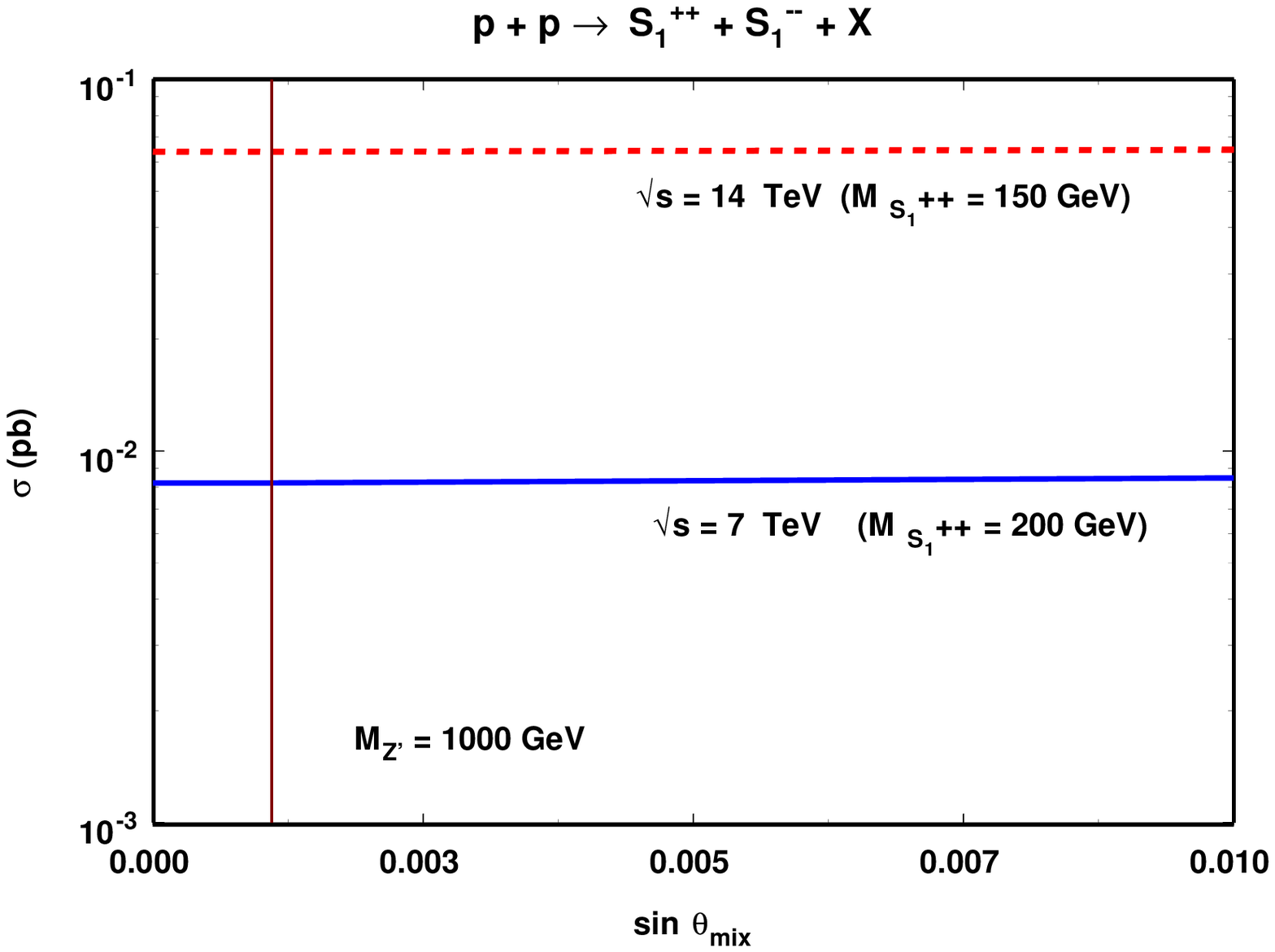}
\caption{The total cross section for $ p + p \lra Y^{++} + Y^{--} + X $ as a function of $\sin \theta_{mix}$ (top). The total cross section for $ p +  p \lra S_1^{++} + S_1^{--} + X $ as 
a function of $\sin \theta_{mix}$ (bottom). Both cross sections where calculated for $\sqrt s = 7$ TeV and $\sqrt s = 14$ TeV. The vertical red line represents the experimental upper bound on $\sin \theta_{mix}$. }
\end{center}
\end{figure}

\vskip 1cm
\textit{Acknowledgments:} 
E. Ramirez Barreto thanks Capes-PNPD. J. S\'a Borges and Y. A. Coutinho thank FAPERJ for financial support.

\ed